# Accelerated alpha-decay of $^{232}$U isotope achieved by exposure of its aqueous solution with gold nanoparticles to laser radiation

**A.V. Simakin, G.A. Shafeev**

Wave Research Center of A.M. Prokhorov General Physics Institute of the Russian Academy of Sciences, Vavilov Street, Moscow 119991, Russian Federation

**Abstract**

Experimental results are presented on laser-induced accelerated alpha-decay of Uranium-232 nuclei under laser exposure of Au nanoparticles in aqueous solutions of its salt. It is demonstrated that the decrease of alpha-activity strongly depends on the peak intensity of the laser radiation in the liquid and is highest at several terawatt per square centimeter. The decrease of alpha-activity of the exposed solutions is accompanied by the deviation of gamma-activities of daughter nuclides of Uranium-232 from their equilibrium values. Possible mechanisms of the laser influence on the alpha-activity are discussed on the basis of the amplification of the electric field of laser wave on metallic nanoparticles.

Laser beams at intensity level of $10^{18}$ W/cm$^2$ are capable of inducing nuclear transformations under laser exposure of solid targets in vacuum. This can be achieved at pico- and femtosecond laser pulse durations [1 – 4]. Laser plasma generated by intense laser wave becomes the source of relativistic protons and electrons. Another approach to laser initiation of nuclear processes consists in laser exposure of nanoparticles (NPs) in liquids. Owing to plasmon resonance of electrons in metallic NPs, its cross section can be much higher than the geometric one. This causes its efficient heating by absorption of laser radiation and further ionization of the nanoparticle material. Expansion of this plasma is confined by surrounding vapors of the liquid. If the surrounding liquid contains compounds of unstable elements, such as Thorium or Uranium, then its activity may significantly deviate from its equilibrium values [5-7]. This deviation occurs at the intensity level as small as $10^{12}$-$10^{13}$ W/cm$^2$. Since the decay scheme of these nuclei is a sequence of both alpha- and beta-decays, one may conclude that laser exposure of their solutions in presence of NPs alters both types of decay. The most probable mechanism of laser acceleration of nuclear decays is the perturbation of electronic shells of unstable elements in the field of intense laser wave [8]. This perturbation alters the inter-atomic field in the vicinity of nuclei and thus alters its stability. The role of NPs is the amplification of the field of the laser wave due to plasmon resonance of charge carriers. The amplification factor may amount to $10^4$ - $10^6$ [9-11], so that the laser intensity of $10^{12}$ W/cm$^2$ at the entrance of the solution may be as high as $10^{16}$ – $10^{18}$ W/cm$^2$ in the vicinity of the nanoparticle. Both components of the light wave are amplified, the electric and magnetic one. The role of these components in the acceleration of different types of nuclear decays is different. Theoretical investigations show that strong magnetic field can significantly accelerate permitted beta-decays [12] and can make possible forbidden beta-decays [13,14]. Acceleration of both alpha- and beta-decays was observed in previous studies of laser influence on the activity of $^{238}$U and nuclides of its branching. Uranium-238 decays via alpha-decay (tunneling of alpha-particles through the potential barrier) with half-life of more than $10^9$ yrs, while his daughter nuclides $^{234}$Th and $^{234m}$Pa decays through beta-decay. Real-time measurements of alpha-activity of $^{238}$U are hardly possible in a liquid medium because of short mean free path of alpha-particles. The conclusion on the variation of alpha-activity of $^{238}$U under laser exposure was made on the basis of indirect measurements of gamma-activity of its daughter nuclides, $^{234}$Th and $^{234m}$Pa [15, 16]. The half-life of alpha-decay of $^{238}$U exceeds $10^9$ years. It is expected that the influence of laser exposure on the alpha-activity will be more noticeable in case of a nuclide with shorter time of alpha-decay. $^{232}$U may serve as such model nuclide. This is an artificial isotope with half-life of alpha-decay of 69 years. The sequence of decays of $^{232}$U is as follows:

$$^{232}\mathrm{U} \xrightarrow{\alpha} {}^{228}\mathrm{Th} \xrightarrow{\alpha} {}^{224}\mathrm{Ra} \xrightarrow{\alpha} {}^{220}\mathrm{Rn} \xrightarrow{\alpha} {}^{216}\mathrm{Po} \xrightarrow{\alpha} {}^{212}\mathrm{Pb} \xrightarrow{\beta} {}^{212}\mathrm{Bi} \xrightarrow{\alpha} {}^{208}\mathrm{Tl} \xrightarrow{\beta} {}^{208}\mathrm{Pb}$$

The aim of the present work is elucidation of the laser exposure of nanoparticles in aqueous solution of $^{232}$U salt on its alpha-activity. The real-time gamma-emission of daughter nuclides $^{212}$Pb and $^{208}$Tl can be used for tracing the deviation from equilibrium under laser exposure.

The alpha-activity of the samples before and after laser exposure was measured using Ortec-65195-P alpha-spectrometer with accuracy of ± 5 %. Aqueous solution of $^{232}UO_2(NO_3)_2$ was divided into several parts. The first part with activity of 53.4 Bq/ml was used as the initial non-exposed solution. This value of activity corresponds to 60 pg of $^{232}$U per ml. Laser exposure of the solution was carried out by focusing the laser beam on a bulk gold target placed inside the solution with the help of an aspherical lens. Initially the solution did not contain nanoparticles, the y were generated during the laser exposure of the target and dispersed in the solution due to convective motion. This scheme provides the constant level of nanoparticles density in the solution since they may agglomerate and sediment during laser exposure.

Four laser sources with different wavelengths and pulse duration were used for the exposure of solutions.

1 – A Nd:YAG laser with wavelength of 1064 nm, pulse duration of 150 ps, estimated peak power in the solution of $10^{13}$ W/cm$^2$. Repetition rate was of 10 Hz, which corresponds to $3.6\times10^4$ laser pulses during 1 hour exposure.

2 – The second harmonics of a Nd:YAG laser with wavelength of 532 nm, pulse duration of 150 ps, estimated peak power in the solution of $10^{12}$ W/cm$^2$. Repetition rate was of 10 Hz, which corresponds to $3.6\times10^4$ laser pulses during 1 hour exposure.

3 – A copper vapor laser with wavelength of 510.6 nm, pulse width of 20 ns, estimated peak power in the medium of $10^{11}$ W/cm$^2$. Repetition rate of this laser was of 15 kHz, which corresponds to $2.16\times10^9$ pulses at 4 hour exposure.

4 – A Nd:YAG laser with wavelength of 1064 nm, pulse duration of 350 ps, estimated peak power in the solution of $10^{12}$ W/cm$^2$. Repetition rate was of 300 Hz, which corresponds to $4.32\times10^7$ laser pulses during 4 hour exposure.

The laser sources have different pulse duration and wavelength of emission. The radiation of both copper vapor laser and the second harmonics of a Nd:YAG laser fits well the position of plasmon resonance of Au nanoparticles in water (around 530 nm). Only approximate estimations of the peak power inside the solution with nanoparticles can be done. This is due to initiation of various non-linear effects inside the liquid at intensity level $10^{11}$ – $10^{13}$ W/cm$^2$ that may lead to defocusing of laser radiation. This can be, for example, the dynamic Kerr effect. On the contrary, self-focusing of the laser beam may lead to its filamentation and to the increase of laser intensity in the medium. Finally, the presence of nanoparticles may alter the intensity level due to the absorption of laser radiation via plasmon resonance of charge carriers in nanoparticles or in the plasma produced around nanoparticles.

Typical alpha-spectrum of the solution is shown in Fig. 1. Alpha-particles that are emitted by $^{232}$U nuclei have higher energy than those emitted by $^{238}$U. This is the reason of much shorter half-life of its alpha-decay.

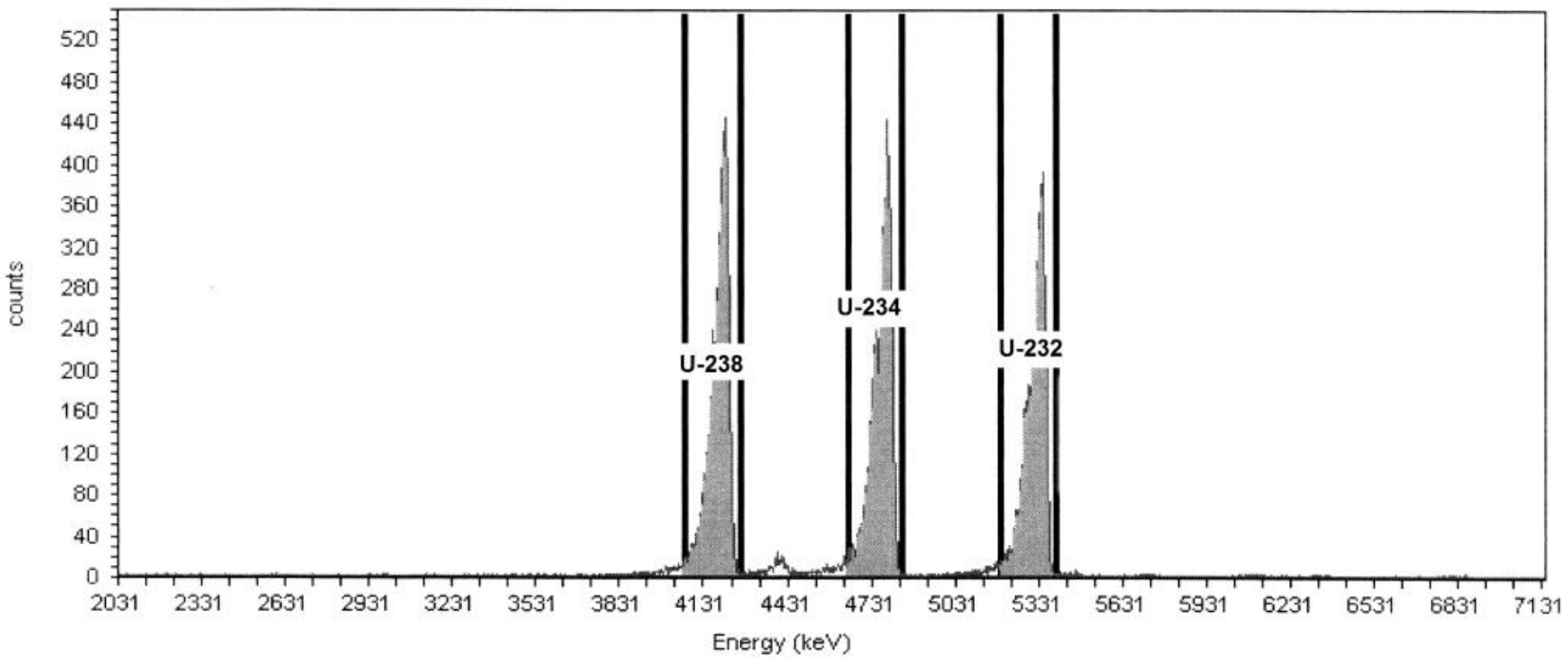

Fig. 1. Alpha-spectrum of the initial solution. Nuclides $^{238}U$ and $^{234}U$ originate from sample solution of $^{238}U$ salt used as a tracer during alpha-measurements.

Change of alpha-activity of the sample produced by exposure of a gold target in aqueous solution of $^{232}U$ salt in water to radiation of a copper vapor laser is accompanied by modification of gamma-activity of its daughter nuclides, such as $^{212}Pb$ and $^{208}Tl$. This is illustrated by the following diagrams:

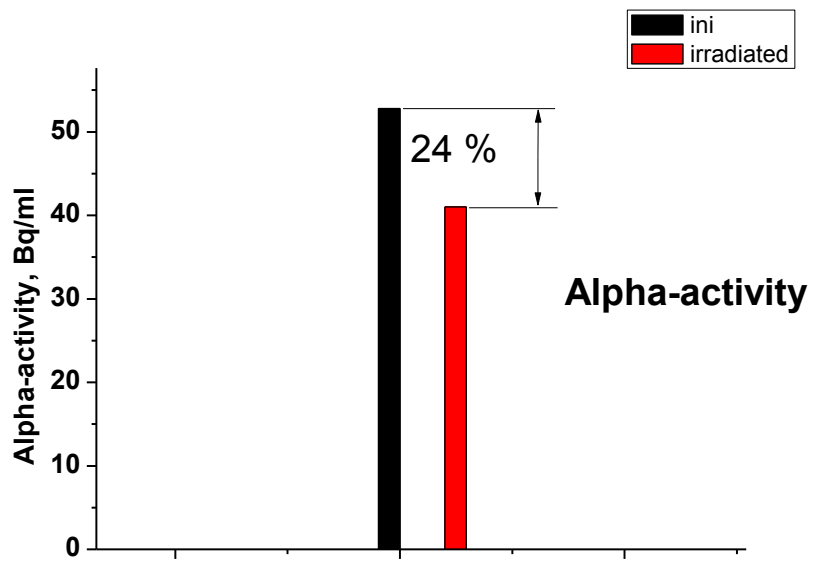

a

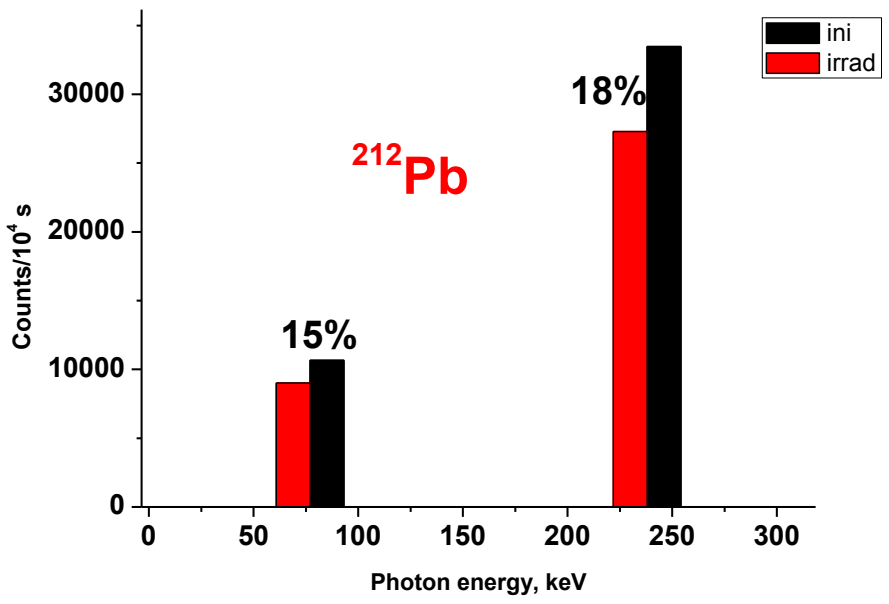

b

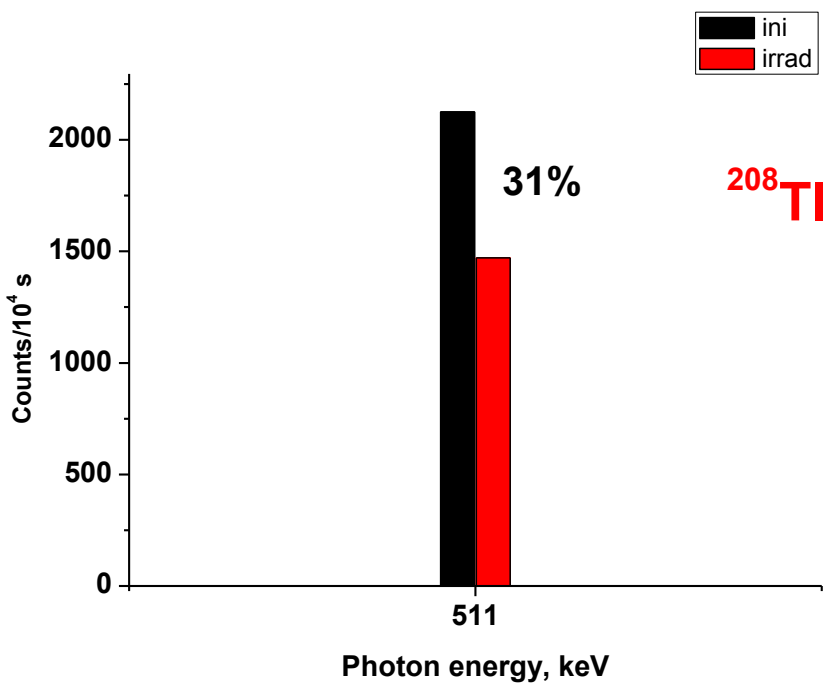

c

Fig. 2. Change of alpha-activity of $^{232}U$ sample (a) and gamma-activities of $^{212}Pb$ (b) and $^{208}Tl$ (c) after exposure of a bulk gold target in aqueous solution of $^{232}UO_2(NO_3)_2$ to radiation of a Cu vapor laser during 4 hours ($2.16\times10^9$ laser pulses).

Real-time gamma-spectroscopy of the solution shows slight distortion of the photon peak of $^{212}Pb$ during laser exposure (Fig. 3).

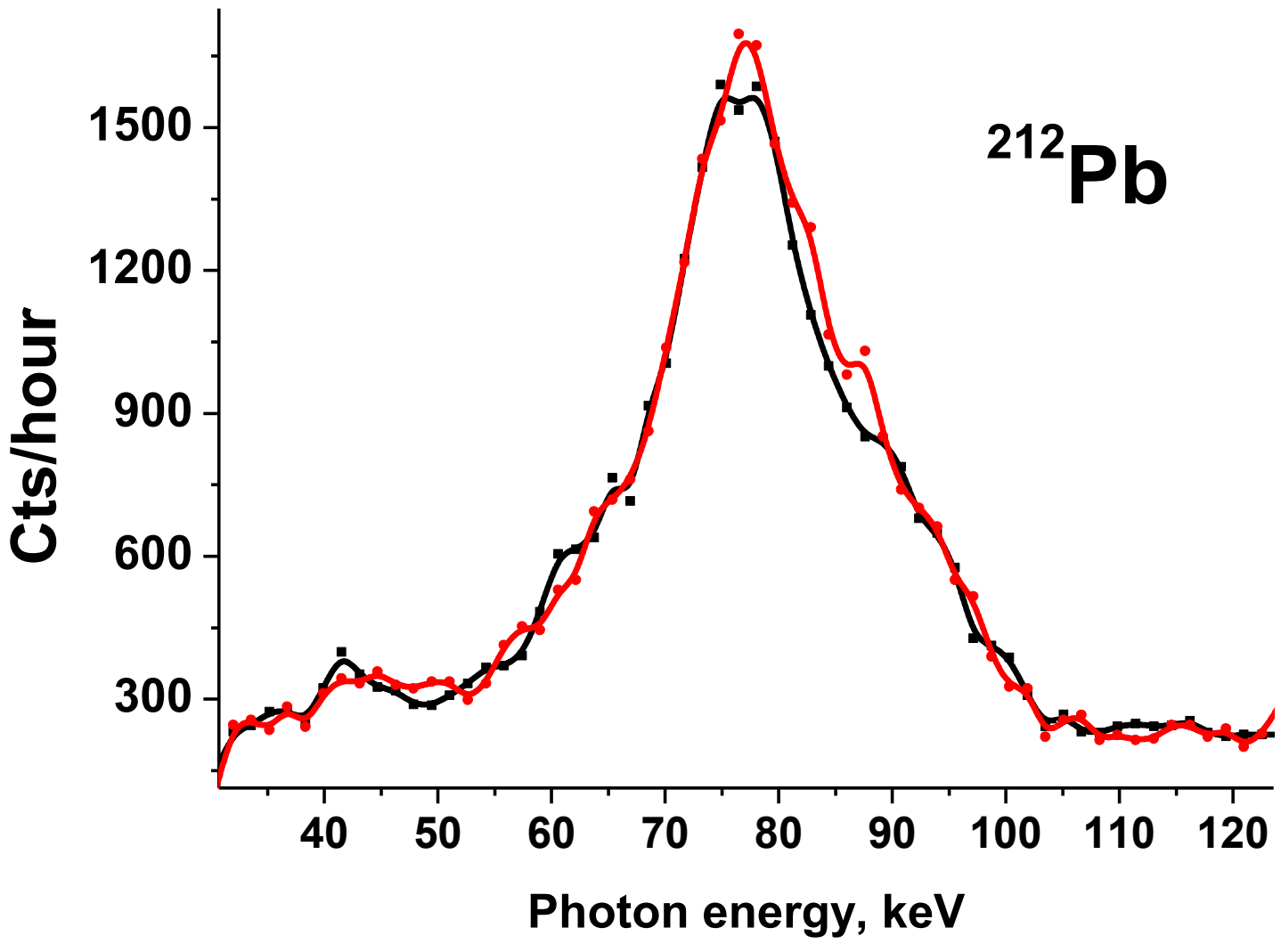

Fig. 3. Real-time profile of 77 keV peak of $^{212}Pb$. Black curve – before laser exposure, red curve – under laser exposure. Cu vapor laser, NaI scintillator gamma-spectrometer (Canberra). Points are channels of the spectrometer.

This observation is in agreement with evolution of non-equilibrium gamma-emission of the solution of $^{238}UO_2Cl_2$ with Au NPs subjected to laser exposure [17]. The main contribution of the gamma-emission of the solution comes from $^{214}$Pb, which is accumulated during accelerated decays of nuclides of $^{238}$U branching.

The effect of exposure of the solution to different laser sources is shown in Fig. 4. One can see that the activity is decreased by a factor of 2 after 1 hour exposure to laser beam at intensity level of $10^{12}$ - $10^{13}$ W/cm$^2$. It is reasonable to suggest that the alpha-decay proceeds during the laser pulse, while the spontaneous decrease of alpha-activity during exposure is negligible. This means that the activity drops down by a factor of 2 during 5 μs, which is the total duration of all 150 ps laser pulses during exposure. In other words, the half-life of $^{232}$U in the laser field is 5 μs instead of 69 years. About $10^{10}$ nuclei of $^{232}$U decay during laser exposure. Another infrared laser radiation with pulse duration of 350 ps affects the alpha-activity of the solution to lesser extent despite to much higher number of laser pulses.

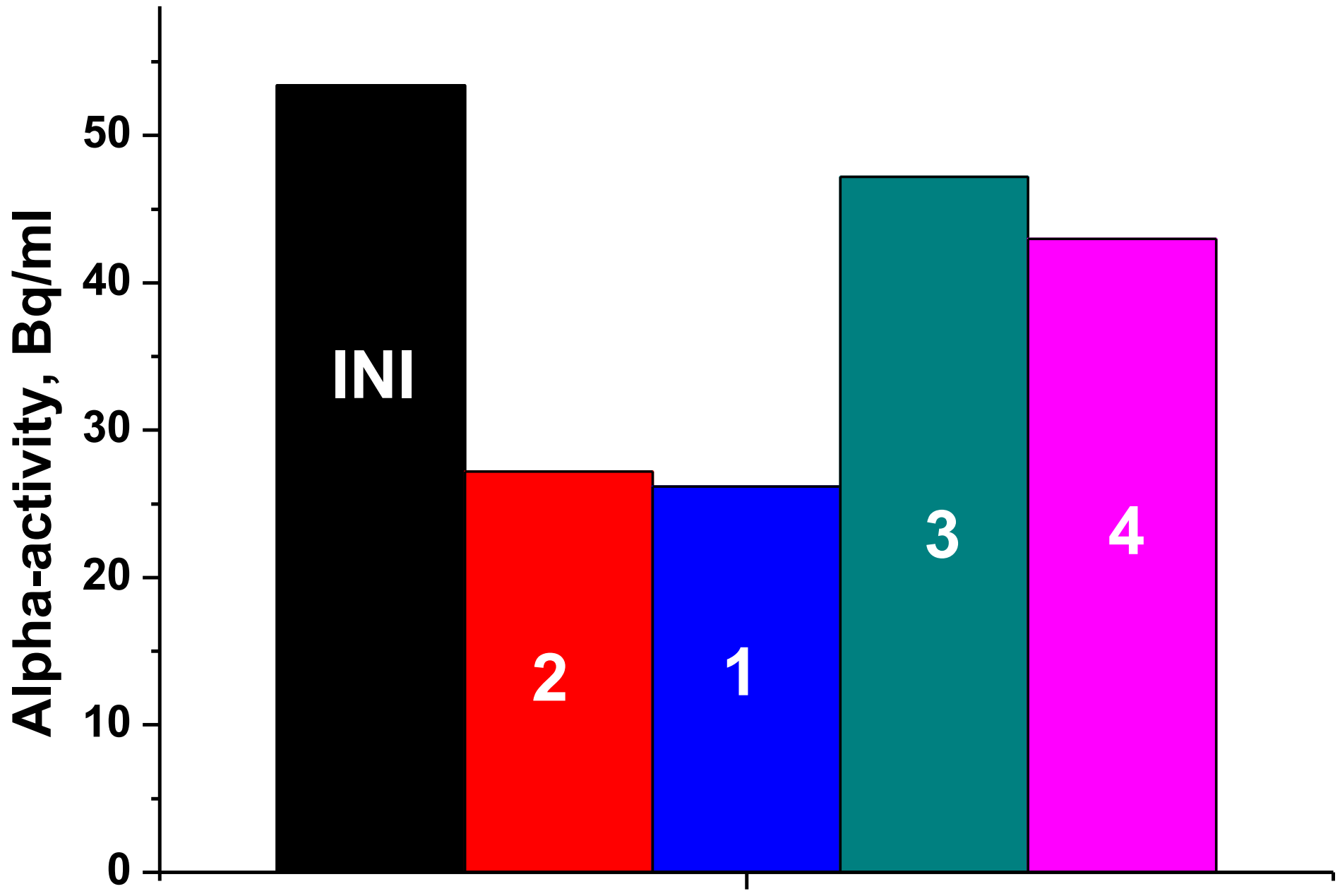

Fig. 4. Diagram of comparison of relative effect of laser exposure on alpha-activity of the same solution to different laser sources. 1 and 2 corresponds to the second and first harmonics of a Nd:YAG laser with pulse duration of 150 ps, respectively. Column 3 indicates the result of exposure to the radiation of a Cu vapor laser. Column 4 stands for the activity after exposure to the first harmonics of a Nd:YAG laser with pulse duration of 350 ps. Initial activity is designated by INI.

Finally, exposure of the solution of $^{232}U$ salt to radiation of a nanosecond Cu vapor laser produces quite small effect on the activity, though this effect is still higher than the accuracy of measurements. This relatively low effect is gained despite to much higher number of laser pulses absorbed in the solution (by a factor of $10^5$) compared to a 150 ps laser source.

The above results one may conclude that the process of laser-induced alpha-decay has no threshold at least down to $10^{11}$ W/cm$^2$.

Strong dependence of the acceleration of alpha-decay on the peak power of laser radiation in the medium should be related to the strength of fields of the laser wave. The natural measure of the electrical field is its value inside the atom or ion. The electric field of laser wave becomes comparable with inter-atomic field at intensity level of $10^{16}$ W/cm$^2$. Possible mechanism of laser-induced acceleration of alpha-decay can be illustrated as follows (Fig. 5). Exposure of NPs to laser radiation leads to its amplification in the vicinity of NPs. If an ion of Uranil is situated near the exposed nanoparticle, then strong electric field of the laser wave disturbes its electronic shells. This perturbation causes the oscillations of the potential near its equilibrium value with the frequency of laser radiation. So do the width and the hieght of the potential barrier for tunneling alpha-particle. Since the probability of tunelling depends on the barrier widt in an exponential way, so even its small variations can noticeably increase the rate of alpha-decay.

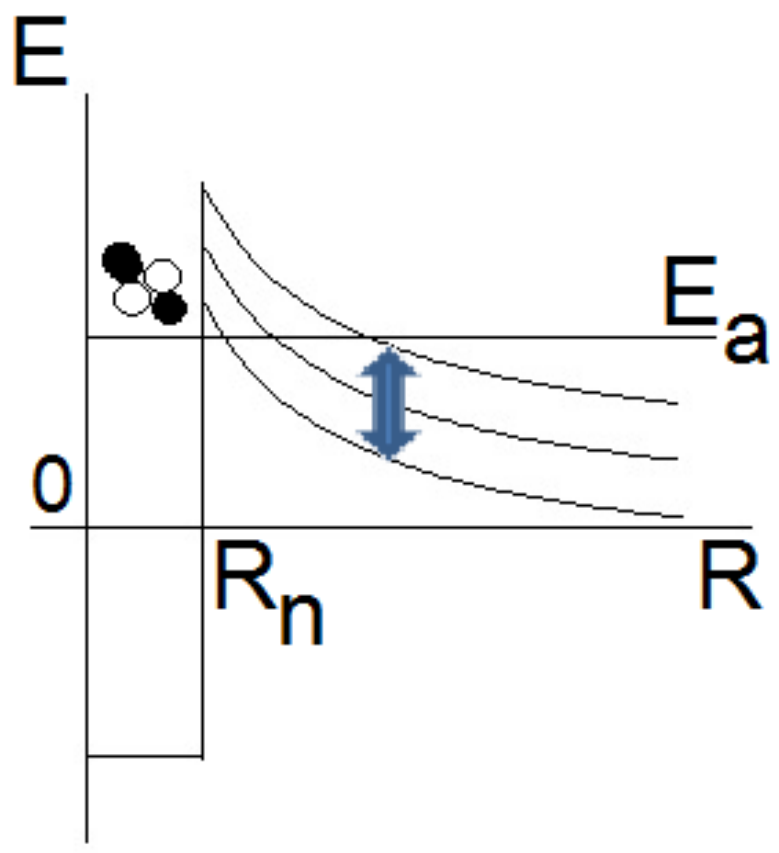

Fig. 5. Qualitative illustration of the possible mechanism of laser-induced acceleration of alpha-decay. Open and closed circles symbolize protons and neutrons of the alpha-particle. $E_a$ stands for the energy of tunneling alpha-particle, $R_n$ is the radius of nuclei.

Owing to high concentration of NPs, the reduction of the width of potential barrier occurs near numerous NPs that are inside the focused laser beam. Also, the effect of laser-induced alpha-decay is observed under numerous laser pulses and has no apparent threshold down to intensities as low as $10^{10}$ W/cm$^2$.

Our recent results show that laser-induced acceleration of alpha-decay can be realized with other isotopes of Uranium. Exposure of the solution of $^{238}UO_2Cl_2$ with Au NPs to radiation of a Nd:YAG laser with pulse duration of 3 ns at estimated peak power of $10^{11}$ W/cm$^2$ during 1 hour at repetition rate of 10 kHz results in the decrease of alpha-activity of $^{238}$U by a factor of 1.6.